\pdfoutput=1
\documentclass[12pt,preprint]{aastex}
\usepackage[margin=0.75in]{geometry}
\usepackage{graphicx}
\usepackage{natbib}
\usepackage{aas_macros}
\usepackage[section] {placeins}
\usepackage{subfigure}
\usepackage{float}
\usepackage{color}

\def\msun{{\rm\,M_\odot}}

\def\msun{{\rm\,M_\odot}} 

\def\zsun{{\rm\,Z_\odot}}

\newcommand{\kms}{\, {\rm km\, s}^{-1}}

\newcommand{\mpc}{\, {\rm Mpc}}
\newcommand{\kpc}{\, {\rm kpc}}
\newcommand{\hmpc}{\, h^{-1} \mpc}

\newcommand{\hi}{\mbox{H\,{\scriptsize I}\ }}
\newcommand{\heii}{\mbox{He\,{\scriptsize II}\ }}

\def\h2{${\rm\,H_2}$}

\makeatletter 

\def\hi{\noindent \hangindent=2.5em}

\def\kpc{{\rm\,kpc}}

\def\mpc{{\rm\,Mpc}}

\def\kms{{\rm\,km/s}}
\def\msun{{\rm\,M_\odot}}

\def\vol#1  {{{#1}{\rm,}\ }}

\def\hi{{H\thinspace I}\ }

\def\heii{{He\thinspace II}\ }

\newcount\refno
\newcount\rfno
\def\eq{$^{\the\refno\ }$\advance\refno by 1}
\def\ad{\advance\rfno by 1}

\def\clock{\count0=\time \divide\count0 by 60
     \count1=\count0 \multiply\count1 by -60 \advance\count1 by \time
     \number\count0:\ifnum\count1<10{0\number\count1}\else\number\count1\fi}

\def\myputfigure#1#2#3#4#5%
{\hskip0.03\textwidth\vskip#5pt
\makebox[0pt]{\hskip#2in
\includegraphics[width=#3\textwidth]{#1}}\vskip#4pt\hfill}

\newcount\refno
\newcount\rfno
\def\eq{$^{\the\refno\ }$\advance\refno by 1}
\def\ad{\advance\rfno by 1}

\definecolor{burntorange}{rgb}{1,0.4,0.2}

\begin{document}

\title{Diverse Properties of Interstellar Medium Embedding Gamma-Ray Bursts at the Epoch of Reionization}
 
\author{
Renyue Cen$^{1}$ and  
Taysun Kimm$^{2}$ 
} 

\footnotetext[1]{Princeton University Observatory, Princeton, NJ 08544;
 cen@astro.princeton.edu}

\footnotetext[2]{Princeton University Observatory, Princeton, NJ 08544;
kimm@astro.princeton.edu}

\begin{abstract} 

Analysis is performed on ultra-high resolution large-scale cosmological radiation-hydrodynamic simulations to,
for the first time, quantify the physical environment of long-duration gamma-ray bursts (GRBs) at the epoch of reionization.
We find that, on parsec scales, 13\% of GRBs remain 
in high density ($\ge 10^4$cm$^{-3}$) low-temperature star-forming regions,
whereas 87\% of GRBs occur in low-density ($\sim 10^{-2.5}$cm$^{-3}$) high temperature regions heated by supernovae.
More importantly, the spectral properties of GRB afterglows, 
such as the neutral hydrogen column density, total hydrogen column density, dust column density, gas temperature and metallicity of intervening absorbers,
vary strongly from sightline to sightline.
Although our model explains extant limited observationally inferred values with respect to circumburst density,
metallicity, column density and dust properties, a substantially larger sample 
of high-z GRB afterglows would be required to facilitate a statistically solid test of the model.
Our findings indicate that any attempt to infer the physical properties (such as metallicity) of the interstellar medium of the host galaxy
based on a very small number of (usually one) sightlines would be precarious.
Utilizing high-z GRBs to probe interstellar medium and intergalactic medium 
should be undertaken properly taking into consideration the physical diversities of the interstellar medium.

\end{abstract}
 

\section{Introduction}

Very high redshift ($z\ge 6$) gamma-ray bursts (GRBs) \citep[e.g.,][]{2009Greiner, 2009Tanvir,2011Cucchiara}
provide an excellent probe of both the interstellar (ISM) and intergalactic medium (IGM) at the epoch of reionization (EoR)
using absorption spectrum techniques thanks to their simple power-law afterglow spectra and high luminosity \citep[][]{2000Lamb},
complimentary to quasar absorption spectrum observations \citep[][]{2006Fan}. 
Here we present a first, detailed analysis of the physical properties of ISM surrounding GRBs,
utilizing state-of-the-art radiation-hydrodynamic simulations, with the hope
that they may aid in proper interpretations of observations of GRB afterglows at EoR with respect to both ISM and IGM.

\section{Simulations}\label{sec: sims}

The simulations are performed using the Eulerian adaptive mesh refinement code, 
{\sc ramses} \citep[][ver. 3.07]{teyssier02}, with concurrent multi-group radiative transfer (RT) calculation \citep{rosdahl13}.
The reader is referred to \citet[][]{2014Kimm} for details. 
Notably, a new treatment of supernova feedback is implemented,
which is shown to capture the Sedov solution for all phases (from early free expansion to late snowplow). 
The initial condition for the cosmological simulations is generated using the {\sc MUSIC} software \citep{2011Hahn}, 
with the WMAP7 parameters \citep{2011Komatsu}:
$(\Omega_{\rm m}, \Omega_{\Lambda}, \Omega_{\rm b}, h, \sigma_8, n_s  = 0.272, 0.728, 0.045, 0.702, 0.82, 0.96)$. 
The total simulated volume of $(25 {\rm Mpc/h})^3$ (comoving) is covered with $256^3$ root grids, 
and 3 more levels are added to a rectangular region of $3.8\times4.8\times9.6$ Mpc to achieve 
a high dark matter mass resolution of $m_{\rm dm}=1.6\times10^5\,M_{\odot}$. 
In the zoomed-in region, cells are further refined (12 more levels) based on the density and 
mass enclosed within a cell. The corresponding maximum spatial resolution of the simulation 
is 4.2 pc (physical). The simulation is found to be consistent with a variety of observations, 
including the luminosity function at $z\sim 7$.

Normal and runaway star particles are created in convergent flows 
with a local hydrogen number density $n_{\rm th} \ge100\,{\rm cm^{-3}}$ \citep[FRU run,][]{2014Kimm}, 
based on the Schmidt law \citep[][]{schmidt59}. Note that the threshold is motivated by the density 
of a Larson-Penston profile \citep{larson69,penston69} at $0.5\Delta x_{\rm min}$, $\rho_{\rm LP}\approx8.86 c_s^2 
/ \pi\,G\,\Delta x_{\rm min}^2$, where $c_s$ is the sound speed at the typical temperature of the ISM ($\sim 30 K$) 
and $\Delta x_{\rm min}$ is the finest cell resolution.
Additionally, we ensure that the gas is Jeans unstable, and that the cooling time is shorter than the dynamical time 
\citep[e.g.][]{cen92}. We assume that  2\% of the star-forming gas is converted into stars per its free-fall time 
($t_{\rm ff}$) \citep{krumholz07}.  The mass of each star particle is determined as
 $m_\star=\alpha\, N_p \rho_{\rm th} \, \Delta x_{\rm min}^3 $, where $\rho_{\rm th}$ is the threshold density,
 and $\alpha$ is a parameter that controls the minimum mass of a star particle ($m_{\rm \star,min}$). 
 $N_p$ is an integer multiple of $m_{\rm \star,min}$ to be formed in a cell, 
 which is drawn from a Poisson random distribution,
$P(N_p) = (\lambda ^{N_p} / N_p! ) \exp\left(-\lambda\right)$ 
with the Poissonian mean 
$\lambda \equiv \epsilon_\star \left(\frac{\rho\Delta x^3}{m_{\rm \star,min}}\right) \left( \frac{\Delta t_{\rm sim}}{t_{\rm ff}}\right)$, 
where $\Delta t_{\rm sim}$ is the simulation time step. 
The resulting minimum mass of a normal (runaway) star particle is $34.2\, M_{\odot} (14.6\, M_{\odot})$. 
We adopt the 
Chabrier 
initial mass function to compute 
the mean frequency of Type II supernova explosions per solar mass ($0.02 M_{\odot}^{-1}$).
Dark matter halos are identified using the Amiga halo finder \citep{2009Knollmann}.
This yields 731 halos of mass $10^{8}\le M_{\rm vir} < 3\times10^{10}\,M_{\odot}$ at $z=7$.
We adopt the assumption that long-duration GRB rate is proportional to type II supernova rate;
short-duration GRBs are not addressed here since they appear not to be associated with massive stars
and are hosted by elliptical galaxies \citep[][]{2013Berger}.
For our analysis we use all snapshots between $z=7.5$ and $z=7$.

\section{Results}

We present results from our simulations and make comparisons to available observations.
As will be clear later, we generally find broad agreement between our model and observations,
although larger observational samples of GRB afterglows would be needed to 
fully test the model.

\subsection{Properties of Embedding Interstellar Medium on Parsec Scales}

\begin{figure}[!h]
\centering
\vskip -0.0cm
\resizebox{3.3in}{!}{\includegraphics[angle=0]{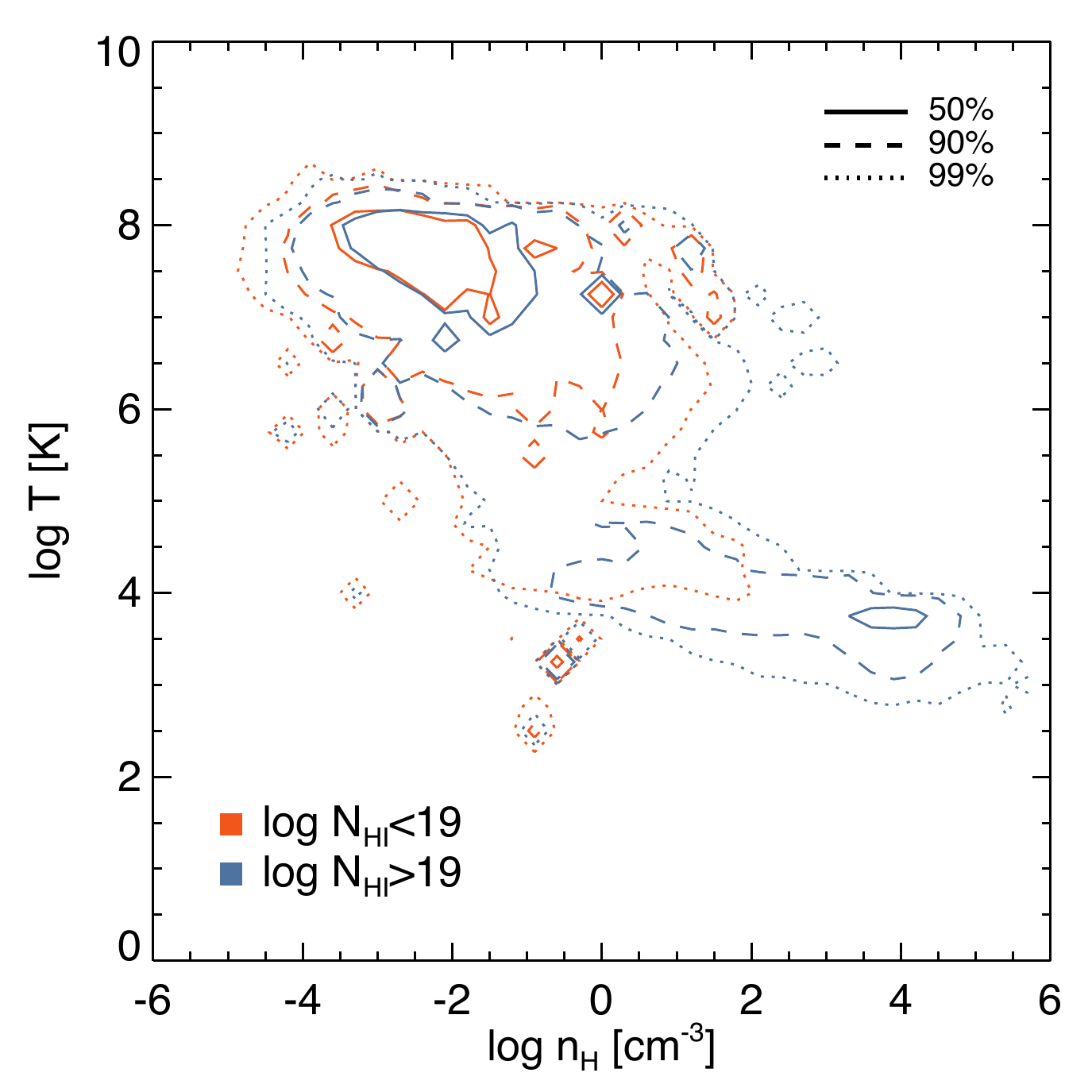}} 
\hskip -0.0cm
\resizebox{3.3in}{!}{\includegraphics[angle=0]{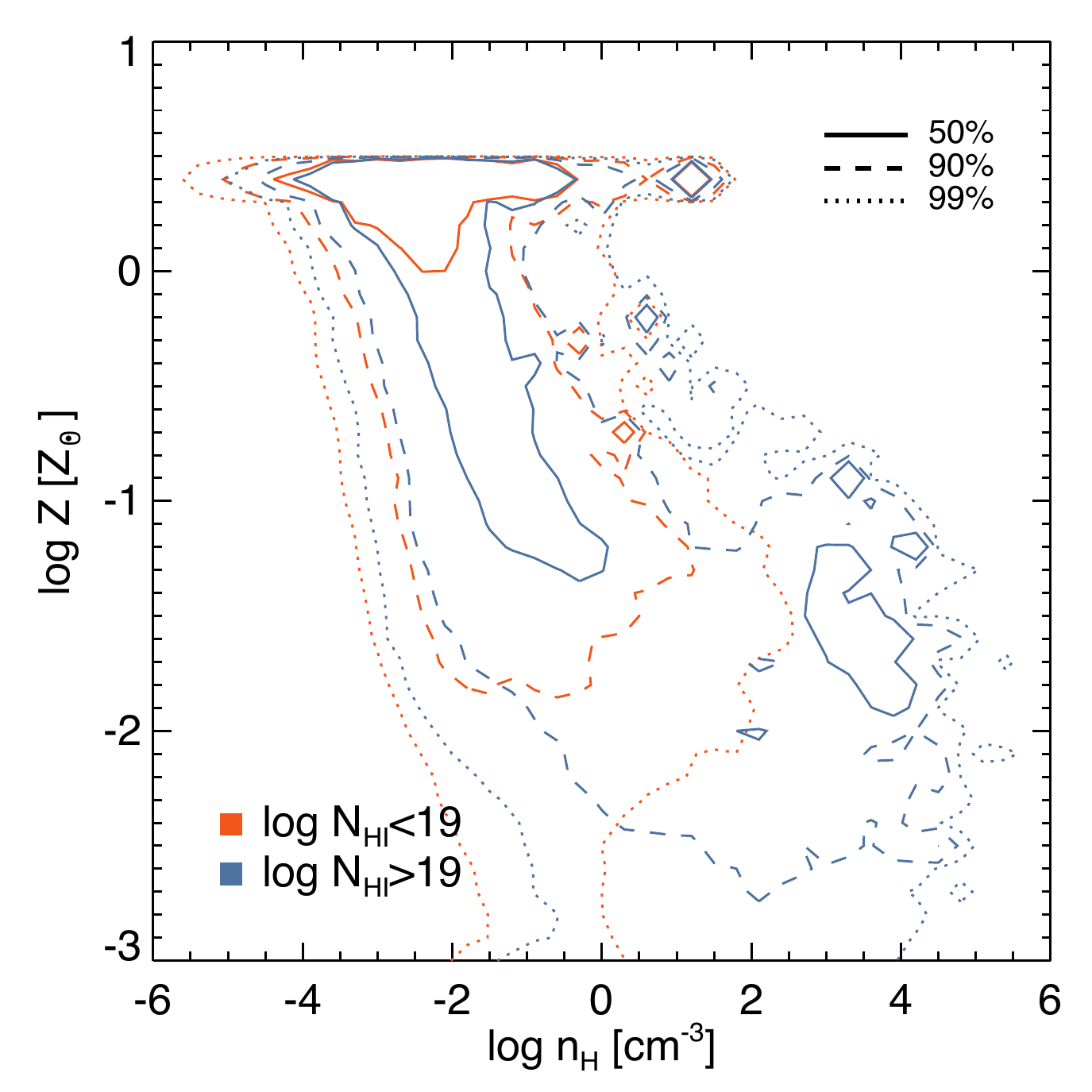}} 
\vskip -0.0cm
\caption{
{\color{burntorange}\bf Left panel:}
shows the distribution of GRB rate in the density-temperature ($n-T$) parameter space.
Note that the density and temperature are defined 
on the local gas cell of scale of a few pc that a GRB sits in and it will be made clear later that
the appearance of GRB afterglows is in most cases more dependent on the properties
of gas along the line of sight.
We have further divided the GRBs into two groups 
with respect to intervening neutral hydrogen column density:
$N_{HI}>10^{19}$cm$^{-2}$ (red)  and $N_{HI}<10^{19}$cm$^{-2}$ (blue), 
details of which will be given in subsequent figures.
The contour levels specified indicate the fraction of GRBs enclosed.
{\color{burntorange}\bf Right panel:}
shows the distribution of GRB rate in the density-metallicity ($n-Z$) parameter space.
}
\label{fig:NT}
\end{figure}

We first describe the physical conditions of the ISM that embeds GRBs on pc scales.
The left (right) panel of Figure~\ref{fig:NT} 
shows the distribution of GRB rate in the density-temperature (density-metallicity) parameter space.
We see two separate concentrations of GRBs in the $n-T$ plane,
with $(n_H,T)$ equal to $(10^{-2.5}{\rm cm}^{-3},10^{7.5}{\rm K})$ and $(10^{4.0}{\rm cm}^{-3},10^{3.8}{\rm K})$,
respectively.
It must be made clear that the density and temperature are defined 
on the local gas cell of scale of a few pc that a GRB sits.
The appearance of GRB afterglow spectra depends, in most case,  more strongly on the properties
of gas along the line of sight rather than the gas immediately embedding them, as will be shown later.
It is seen that most of the GRBs reside in the low density, high temperature peak,
contributing to 87\% of GRBs.
It is easy to identify two corresponding concentrations in the $Z-n$ plane in the right panel:
the low density, high temperature peak 
corresponds the high metallicity peak in the range $[-1.5, 0.5]$ in solar units,
while the high density, low temperature peak 
corresponds the low metallicity peak in the range $[-2, -1]$.
We note that super solar metallicities in hot winds driven by type II supernova explosions
in starburst galaxies in conditions similar to those of our simulated galaxies are locally observed.
For example, \citet[][]{2011Konami} observe metallicity of hot X-ray gas of $2-3$ times the solar value
in M82. 
\citet[][]{2002Martin} find that the best fit model for the hot X-ray gas metallicity 
in dwarf starburst galaxy NGC 1569 is solar, although a metallicity as high as 5 times solar still gives
$\chi^2$ value only about 0.1\% larger than the best fit model;
on the contrary, the model with 0.25 times solar has much worse $\chi^2$ value.
The extant observations of GRB afterglow spectra
do not have the capability to detect the metallicity of the X-ray absorbing medium of relatively low column density.
Metallicities of lower temperature gas phases are observed and 
are predicted to be substantially sub solar, as will be shown in 
Figure~\ref{fig:Zg} later.

The $(n_H,T)=(10^{4.0}{\rm cm}^{-3},10^{3.8}{\rm K})$ peak coincides with the 
cores of dense gas clouds in the simulation, where star formation is centered.
The temperature of $10^{3.8}{\rm K}$ is mostly produced by atomic hydrogen cooling and the lower temperature extension
due to metal cooling included in the simulation.
As a numerical example, for a gas parcel of density $10^4$cm$^{-3}$, temperature of $10^4$K and metallicity of 1\% of solar value,
the cooling time is about $400$~yrs, which is much shorter than relevant dynamic time scales.
It is thus clear that the cold density phase seen our simulation is easily understandable.
However, due to lack of treatment for molecular hydrogen cooling, gas is unable to cool significantly below $\sim 10^4$K.
Had we included molecular hydrogen cooling and low temperature metal cooling, we expect the gas to cool approximately isobarically to about $20$K.
Thus, we prefer not to infer any observable properties of GRBs that would depend strongly on the nature of this cold gas phase, such as molecular clouds. 
It is more appropriate to treat $(n_H,T)=(10^{4.0}{\rm cm}^{-3},10^{3.8}{\rm K})$ as bounds: 
$(n_H,T)=(>10^{4.0}{\rm cm}^{-3},<10^{3.8}{\rm K})$.
This is also why we present our results for the high-density low-temperature regions as bounds in the abstract and conclusions sections
as well as places where clarification is helpful.
The noticeable sub-dominance of GRBs residing in very high density ($n\sim 10^4$cm$^{-3}$), star-forming regions suggests that
a large number of stars are displaced from their birth clouds.
This may be achieved by a substantial relative motions between stars and their birth clouds due to hydrodynamic interactions of the later
or dynamical effects of stars.
As a numerical illustration for the former possibility, a relative motion of $10\kms$ between the birth cloud and the star would 
yield a displacement of $100$pc in a lifetime of $10$Myr.
We note that the runaway OB stars in our simulation have typical velocities relative to the birth clouds
of $20-40\kms$. Thus the runaway OB stars have contributed significantly to the displacing GRBs from their birth clouds.
The GRBs being in hot low density environment is also a result 
of supernova heating by earlier supernovae exploding in the birth clouds.
We estimate that these two effects are responsible about equally for placing most of the GRBs in low-density high temperature regions.
While it is not possible to locate GRBs within the host galaxies at high redshift at this time,
observations of low redshift GRBs may still be instructive.
\citet[][]{2012LeFloch} show that GRB 980425 occurring in a nearby ($z=0.0085$) SBc-type dwarf galaxy 
appears to be displaced from the nearest H II region by $0.9$kpc,
which is in fact significantly larger than the displacement distances for the vast majority of our simulated GRBs in high redshift galaxies.

Interestingly, the optical afterglow luminosity has a bimodal distribution at 12 hours after trigger \citep[][]{2008Nardini}.
The bimodal distribution of volumetric density seen in the left panel of Figure~\ref{fig:NT}
alone should produce a bimodal distribution of the afterglows with respect to 
break frequencies, luminosities, and break times, etc \citep[e.g.,][]{1998Sari}.
We cannot make detailed comparisons, because the circumburst density of the 
high $n_{\rm H}$ GRB subset is underestimated due to our limited resolution 
and because it remains uncertain if the appearance of GRB afterglows would also depend strongly 
on intervening material (dust obscuration, etc).
It is suggestive that the complex situations seen in simulations may account for the 
observed bimodality of afterglows without having to invoke intrinsic bimodality of GRBs.

\subsection{Strong Variations of Intervening Gas and Dust along Different Sightlines}

\begin{figure}[!h]
\centering
\vskip -0.0cm
\resizebox{4.0in}{!}{\includegraphics[angle=0]{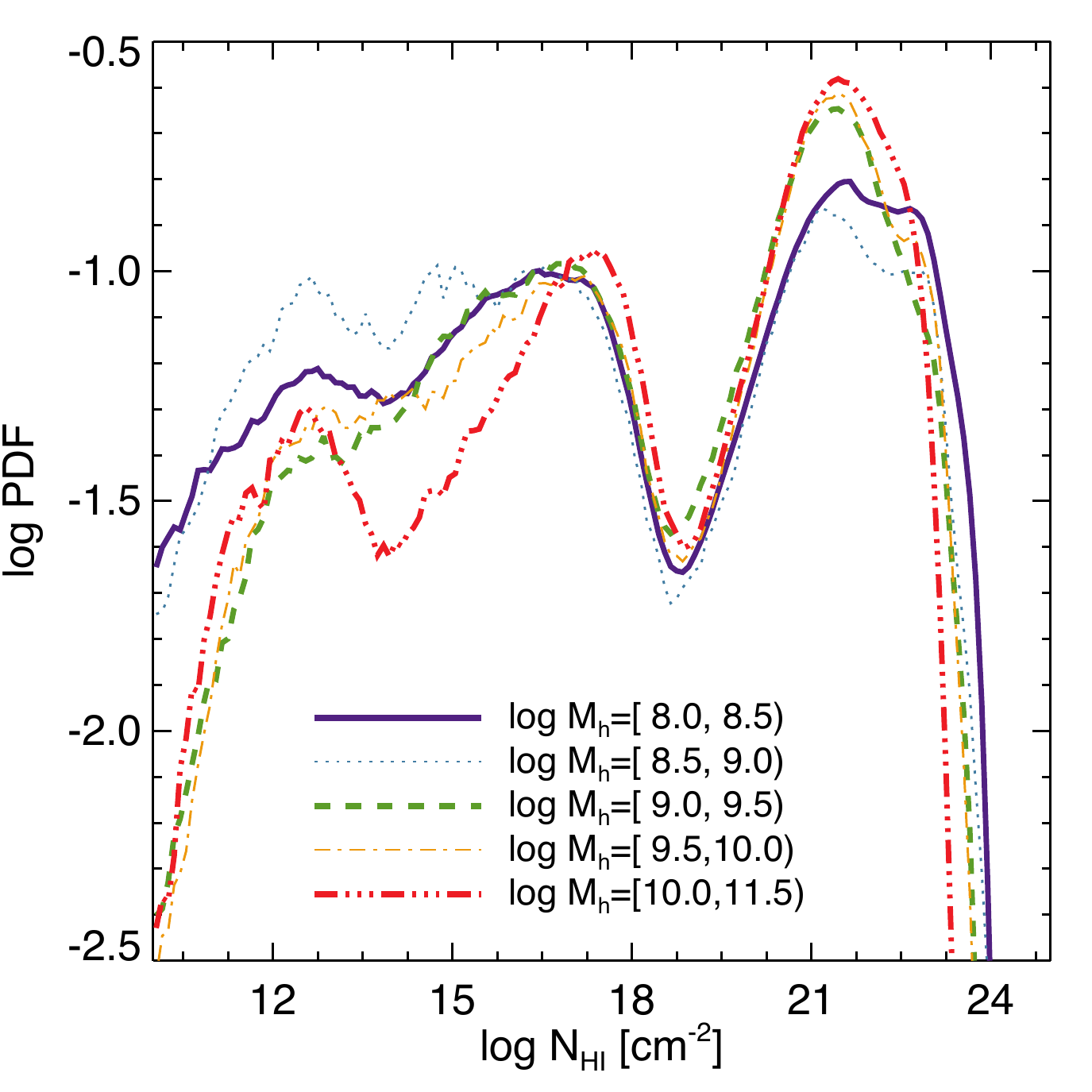}} 
\vskip -0.0cm
\caption{
shows the probability distribution functions (PDF) of neutral hydrogen column density 
for all GRBs, separated according to the halo masses indicated in the legend.
}
\label{fig:NHItoge}
\end{figure}

One of the most important points that this paper hopes to highlight and convey 
is that the appearances of the afterglows of GRBs
are not solely determined by the circumburst medium in their immediate vicinity 
(e.g., the physical conditions shown in Figure~\ref{fig:NT} are on pc scales centered on GRBs).
They also strongly depend on the line of sight
beyond the immediate circumburst medium through the ISM in the host galaxy, which we now quantify.
Let us first give the meaning for our chosen value of the intervening neutral hydrogen column density $N_{HI}=10^{19}$cm$^{-2}$,
which is used  in Figure~\ref{fig:NT} to separate GRBs into separate groups.
Figure~\ref{fig:NHItoge} shows
the distribution of neutral hydrogen column density integrated along the line of sight for all GRBs, 
separately for halos in five mass ranges.
A bimodal distribution of $N_{\rm HI}$ is seen, 
peaked at $N_{\rm HI}\sim 10^{21-22}$cm$^{-2}$ and $N_{\rm HI}\sim 10^{16-17}$cm$^{-2}$, respectively,
well separated by $N_{\rm HI}\sim 10^{19}$cm$^{-2}$.
It is clear that the bimodality exists for all halo masses surveyed.
The low $N_{\rm HI}$ peak is rather broad, extending all the way to $N_{\rm HI}=10^{11}$cm$^{-2}$,
suggesting some locations of GRBs well into the diffused hot ISM.
There is a noticeable dip in the neutral hydrogen column density distribution at $\sim 10^{14}$cm$^{-2}$ for the most massive galaxies of $\ge 10^{10}\msun$.
We attribute this to more significant shock heating in the most massive halos.

Returning to Figure~\ref{fig:NT}, 
it is now easy to see that the GRBs in the low $N_{\rm HI}$ ($\le 10^{19}$cm$^{-2}$) 
peak in Figure~\ref{fig:NHItoge} 
is composed of only one set of GRBs situated in low density environment around 
$(10^{-2.5}{\rm cm}^{-3},10^{7.5}{\rm K})$, seen as the red contours in Figure~\ref{fig:NT}. 
The high $N_{\rm HI}$ ($\ge 10^{19}$cm$^{-2}$)  
peak in Figure~\ref{fig:NHItoge}, on the other hand,
consists of a combination of two separate populations 
with distinctly different circumburst medium,
which correspond to two separate loci of the blue contours 
at $(10^{-2.5}{\rm cm}^{-3},10^{7.5}{\rm K})$ and $(10^{4.0}{\rm cm}^{-3},10^{3.8}{\rm K})$ 
in Figure~\ref{fig:NT}.

The apparently two different groups of GRBs situated around $(n_H,T)=(10^{-2.5}{\rm cm}^{-3},10^{7.5}{\rm K})$
- one with low $N_{\rm HI}$$\le 10^{19}$cm$^{-2}$ (red) 
and the other with low $N_{\rm HI}$$\ge 10^{19}$cm$^{-2}$ (blue) -
are due entirely to the line of sight through the ISM of the host galaxy.
Overall, we find that 38\% of GRBs have $N_{\rm HI}$$\le 10^{17}$cm$^{-2}$ (i.e., optically thin to Lyman continuum),
whereas 44\% have  $N_{\rm HI}$$\ge 10^{20.3}$cm$^{-2}$ (i.e, containing a damped Lyman-alpha system).
It is clear that various properties of the afterglows of GRBs, even sitting in the same very local environment on pc scales,
may appear different due to different intervening interstellar gas and dust along the line of sight through the host galaxy.
In summary so far, there are three separate populations of GRB afterglows are expected, if our model is correct.
One might classify them in the following simple way:
(1) HnHN=(high volumetric density $n\sim 10^4$cm$^{-3}$, high neutral column density $N_{\rm HI}$$\ge 10^{19}$cm$^{-2}$), 
(2) LnHN=(low volumetric density $n\sim 10^{-2.5}$cm$^{-3}$, high neutral column density $N_{\rm HI}$$\ge 10^{19}$cm$^{-2}$), 
(3) LnLN=(low volumetric density $n\sim 10^{-2.5}$cm$^{-3}$, high neutral column density $N_{\rm HI}$$\le 10^{19}$cm$^{-2}$). 
Again, types (2) and (3) are a result of different viewing angles, where type (2) is due to viewing angles
through largely hot ionized gas and type (3) viewing angles going through cold and dense gas in addition.

\citet[][]{2014Laskar} analyze multi-wavelength observations of the afterglow of GRB 120521C ($z\sim 6$)
and re-analyze two previous GRBs at $z>6$ (GRB 050904 and 090423),
and conclude that the circumburst medium has a volumetric density
of $n_H\le 0.05$cm$^{-3}$ that is constant.
The GRBs in the LnHN or LnLN group provide the right match to the observations.
While the statistic is still small, it is expected that about 87\% of GRBs should arise in 
in either the LnHN or LnLN group.

Observations of GRB 050904 at $z=6.3$ reveal that it contains a damped Lyman alpha systems (DLAs) system in the host galaxy of
column density $N_{HI}=10^{21.6}$cm$^{-2}$ and metallicity of $Z=-2.6$ to $-1$
\citep[][]{2006Totani, 2006Kawai}.
Based on X-ray observations, \citet[][]{2007Campana} conclude that $Z\ge 0.03\zsun$ for GRB 050904. 
The evidence thus suggests that GRB 050904 likely resides in a dense environment, although it cannot
be completely sure because the metallicity range of the low-density peak (right panel of Figure~\ref{fig:NT})
overlaps with the observed range.
It is useful at this juncture to distinguish between the metallicity of the local environment of a GRB
and that of absorbers in the GRB afterglow spectrum.

\begin{figure}[!h]
\centering
\vskip -0.0cm
\resizebox{4.0in}{!}{\includegraphics[angle=0]{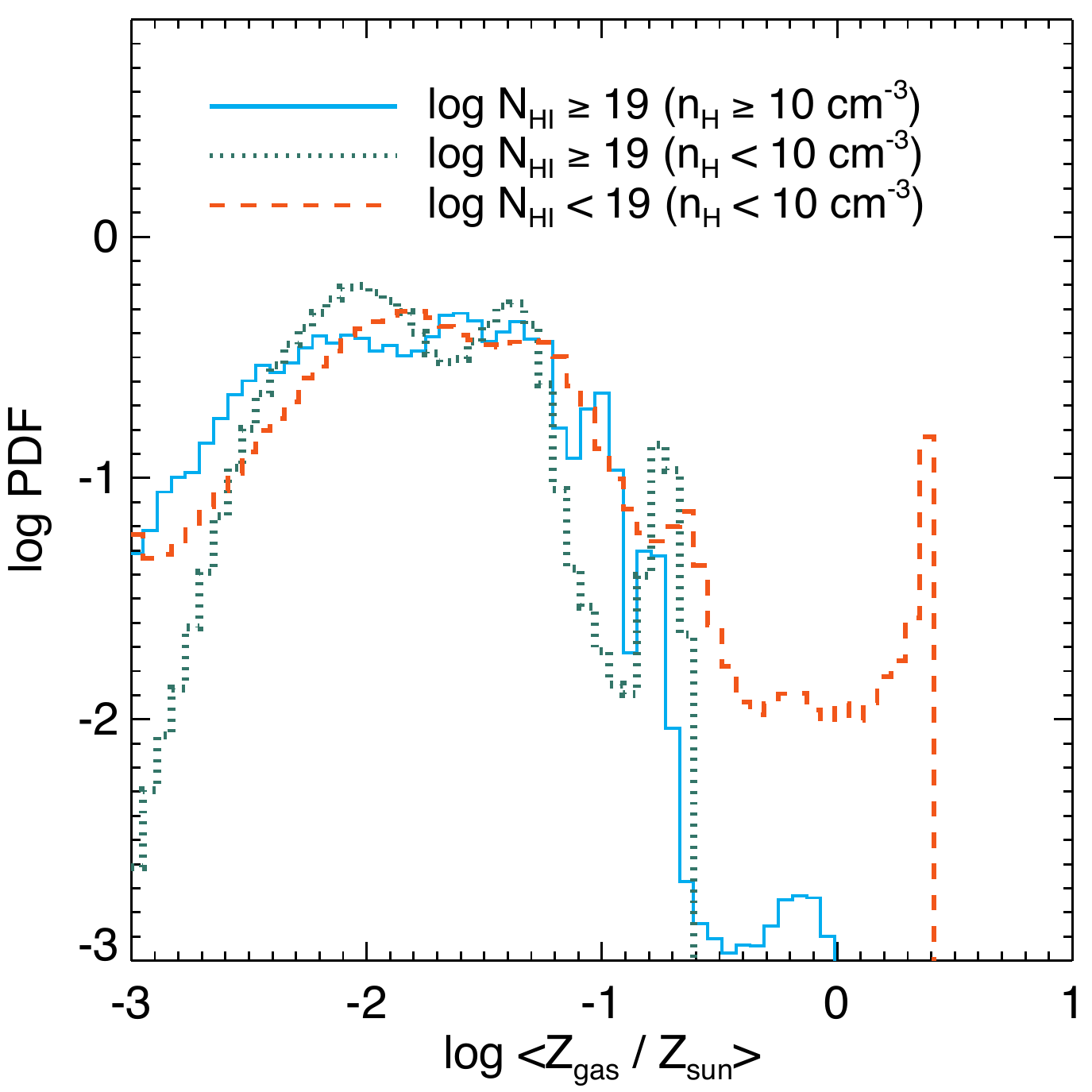}} 
\vskip -0.0cm
\caption{
shows the PDFs of total hydrogen column density weighted metallicity of gas along
the line of sight, excluding gas hotter than $10^6$K,
for the three sub-populations of GRBs:
GRBs in $(n_H,T)=(10^{-2.5}{\rm cm}^{-3},10^{7.5}{\rm K})$ with $N_{\rm HI}\le 10^{19}$cm$^{-2}$ (red dashed, LnLN group),
$(n_H,T)=(10^{-2.5}{\rm cm}^{-3},10^{7.5}{\rm K})$ with $N_{\rm HI}\ge 10^{19}$cm$^{-2}$ (green dotted, LnHN group),
and 
$(n_H,T)=(10^{4.0}{\rm cm}^{-3},10^{3.75}{\rm K})$ with $N_{\rm HI}\ge 10^{19}$cm$^{-2}$ (blue solid, HnHN group).
}
\label{fig:Zg}
\end{figure}

Let us now turn to the expected metallicity of UV/optical absorbers in the GRB afterflow spectra.
Figure~\ref{fig:Zg} shows the PDFs of total hydrogen-column-density-weighted metallicity of gas along
the line of sight, excluding gas hotter than $10^6$K, for the three sub-populations of GRBs.
We see that for all three GRB groups the metallicity of the absorbers in the GRB spectra
peaks in the range $-3$ to $-1$. 
Thus, it is now clear that our model can easily account for the observed properties of GRB 050904.
The additional evidence that, based on the analysis of
the equivalent width ratio of the fine structure transition lines Si II* $\lambda$1264.7\AA\  and Si II $\lambda$1260.4\AA,
infers the electron density of $\log n_e = 2.3\pm 0.7$.
Furthermore, the magnitude of the optical afterglow at 3.4 days after the burst favors a high density circumburst medium.
In combination, it appears that GRB 050904 is likely in  a dense environment being to the HnHN group.
This appears to be at some minor odds with our model, since we only expect that 13\% of GRBs to arise in the HnHN group.
It would be highly desirable to obtain a larger sample of high-z GRBs to provide a statistically firmer test.

Analyses of observations of GRB 130606A at $z=5.9$ 
indicate that it likely contains a sub-DLA system of $N_{HI}\sim 10^{19.8}$cm$^{-2}$ in the host galaxy 
\citep[][]{2013Totani, 2013CastroTirado}.
The inferred low metallicity of $-1.8$ to $-0.8$ in solar units 
\citep[][]{2013CastroTirado} and $-1.3$ to $-0.5$ \citep[][]{2013Chornock},
in conjunction with the $N_{\rm HI}$,
suggests that GRB 130606A may reside in a low density environment with
a foreground sub-DLA system in the host galaxy.
This proposal is consistent with the evidence of detection of highly ionized 
species (e.g., N V and Si IV) \citep[][]{2013CastroTirado}.
It seems likely that GRB 130606A belongs to the LnHN group.

It is easy to see that in our model 
the metallicity distribution of UV/optical absorbers in GRB afterglow spectra is wide,
which itself is due to the very inhomegeneous metallicity distributions in the ISM of the host galaxies.
Thus, it would be a rather chancy practice trying to infer 
the metallicity of the host galaxy solely based on a small number of (typically one) GRB afterglow absorption spectra.

\begin{figure}[!h]
\centering
\vskip 0.5cm
\resizebox{3.0in}{!}{\includegraphics[angle=0]{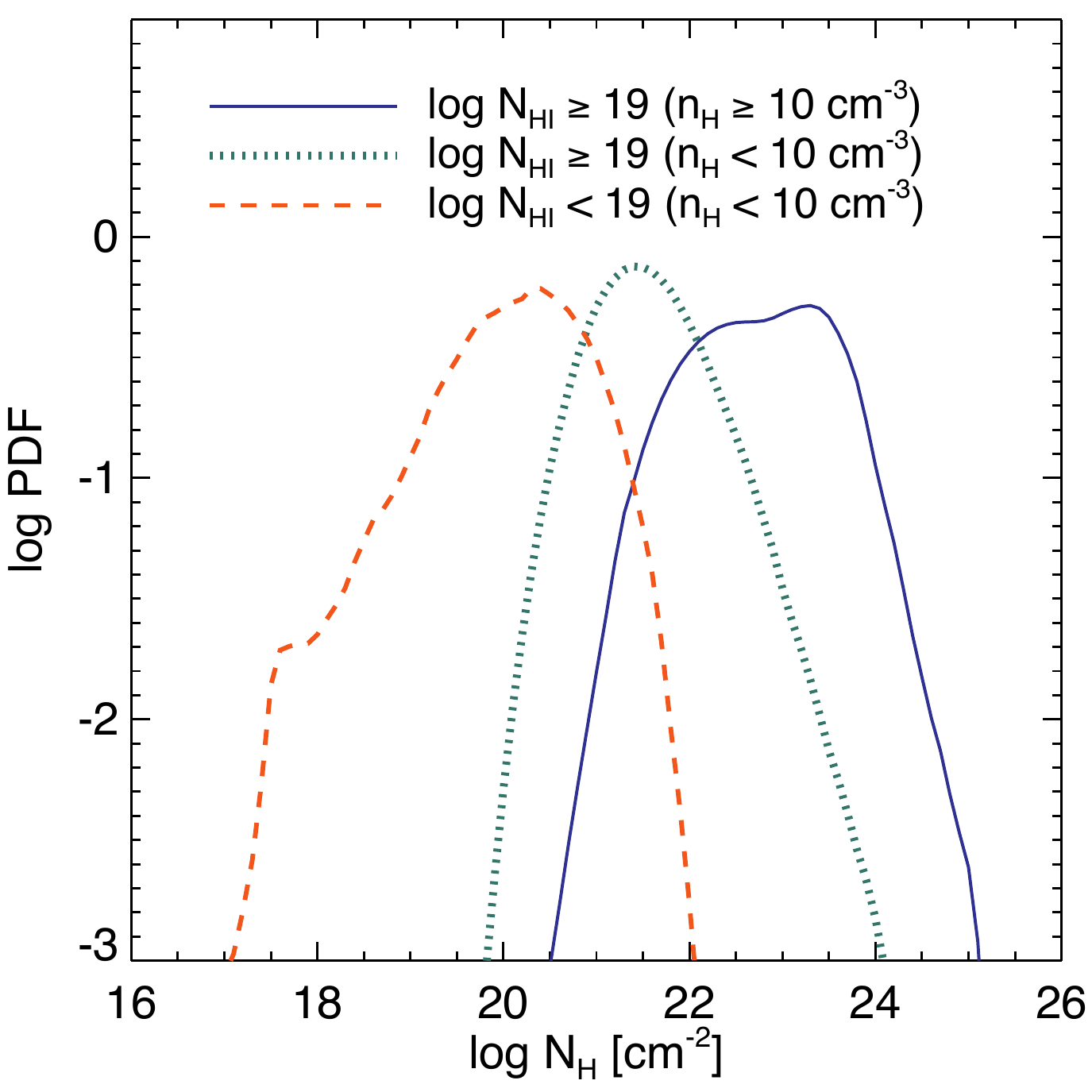}} 
\resizebox{3.0in}{!}{\includegraphics[angle=0]{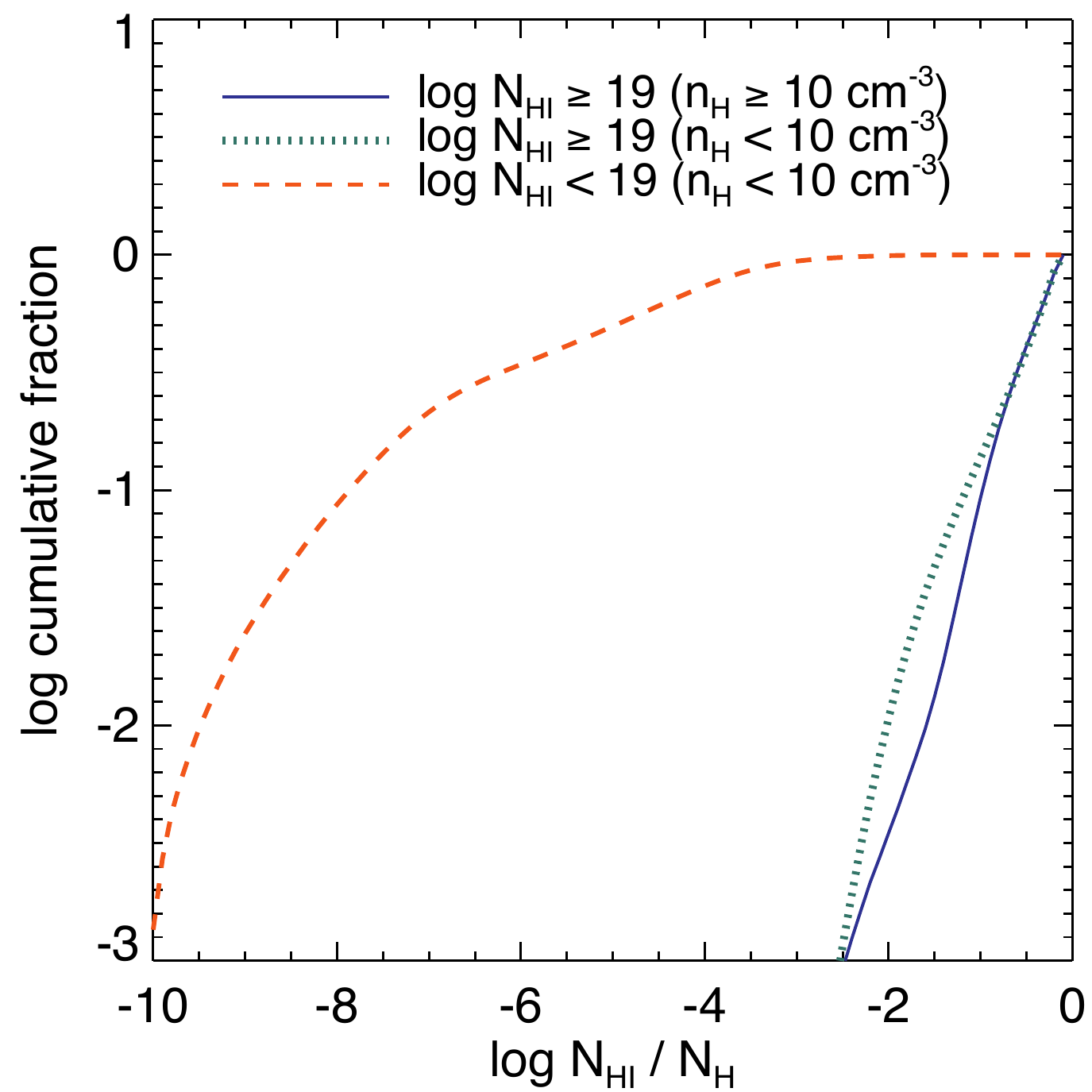}} 
\vskip -0.0cm
\caption{
{\color{burntorange}\bf Left panel:}
shows the PDFs of total hydrogen column density 
for the three sub-populations of GRBs:
GRBs in $(n_H,T)=(10^{-2.5}{\rm cm}^{-3},10^{7.5}{\rm K})$ with $N_{\rm HI}\le 10^{19}$cm$^{-2}$ (red dashed, LnLN group),
$(n_H,T)=(10^{-2.5}{\rm cm}^{-3},10^{7.5}{\rm K})$ with $N_{\rm HI}\ge 10^{19}$cm$^{-2}$ (green dotted, HnLN group),
and 
$(n_H,T)=(10^{4.0}{\rm cm}^{-3},10^{3.75}{\rm K})$ with $N_{\rm HI}\ge 10^{19}$cm$^{-2}$ (blue solid, HnHN group).
{\color{burntorange}\bf Right panel:}
shows the cumulative PDFs of the ratio of $N_{HI}/N_{H}$.
}
\label{fig:NHn}
\end{figure}

The reader has already seen clearly that the distributions of all concerned physical quantities,
including metallicity, density, total and neutral hydrogen column density, are wide.
We will add yet one more quantity and show the cumulative distributions of the ratio of neutral hydrogen to total hydrogen column density for the three groups
in the right panel of Figure~\ref{fig:NHn}. 
We see that for GRBs in the LnLN group the neutral hydrogen to total hydrogen column ratio
is significantly less than unity.
Even for the HnHN and LnHN groups, (10\%,14\%) of GRBs have the ratio less than $0.1$.
In other words, it is generally a pretty bad assumption that the apparent absorbers in the GRB afterglow spectra 
are mainly neutral. This indicates that the so-called ``missing gas problem" 
\citep[e.g.,][]{2011Schady} may be accommodated in this model.

The left panel of Figure~\ref{fig:NHn} shows the PDFs of the total hydrogen column density for the three sub-populations of GRBs,
which is most relevant for probing GRB X-ray afterglows and hence a useful test of our model.
One expectation from our model is that the vast majority of GRBs sitting in low density circumburst medium (LnHN + LnLN) 
do not have Compton thick obscuring gas.
This prediction is verifiable with a combination of afterglow light curves and X-ray observations.
On the other hand, one expects from our model that a significant fraction of the GRBs sitting in high density 
circumburst medium (HnHN) have an extended high $N_{H}$ tail
and dominate the GRBs with $N_{\rm H}\ge 10^{23}$cm$^{-2}$. 
Quantitatively, we find that (45\%, 3\%) of GRBs have $N_{\rm H}\ge (10^{23},10^{24}) $cm$^{-2}$; 
it is noted that these two numbers are likely lower bounds due to possible numerical resolution effects.
As already noted earlier, it is seen from the right panel that
the GRBs in the LnLN group are intervened by highly ionized gas peaking 
at an average neutral fraction of $\sim 10^{-4}$, with no cases having a neutral fraction exceeding $10^{-1}$.
In contrast, for GRBs in both the HnLN and HnHN groups,
more than 50\% of them have an average neutral fraction greater than $\sim 10^{-1}$.

\begin{figure}[!h]
\centering
\vskip 0.5cm
\resizebox{3.3in}{!}{\includegraphics[angle=0]{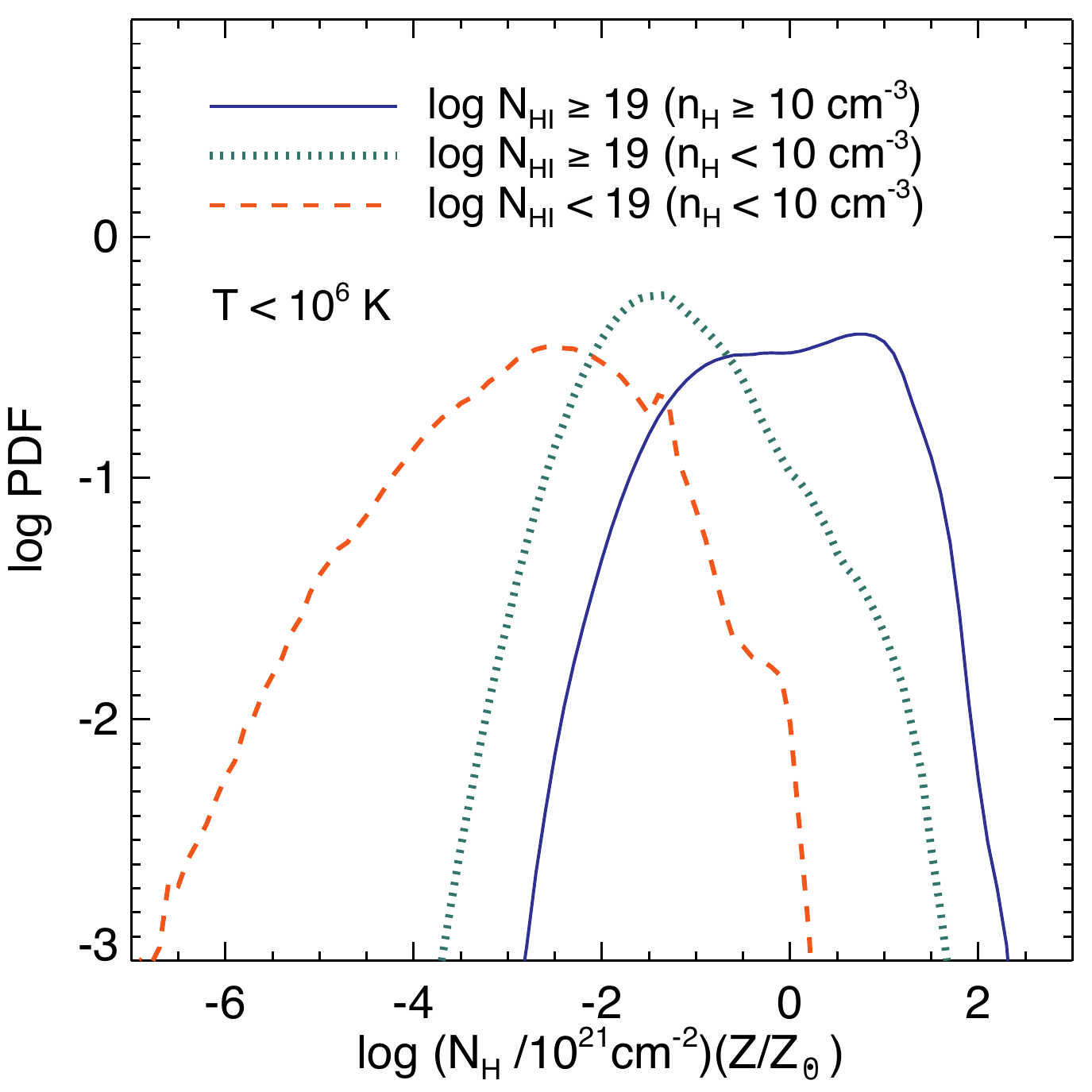}} 
\hskip -0.0cm
\resizebox{3.3in}{!}{\includegraphics[angle=0]{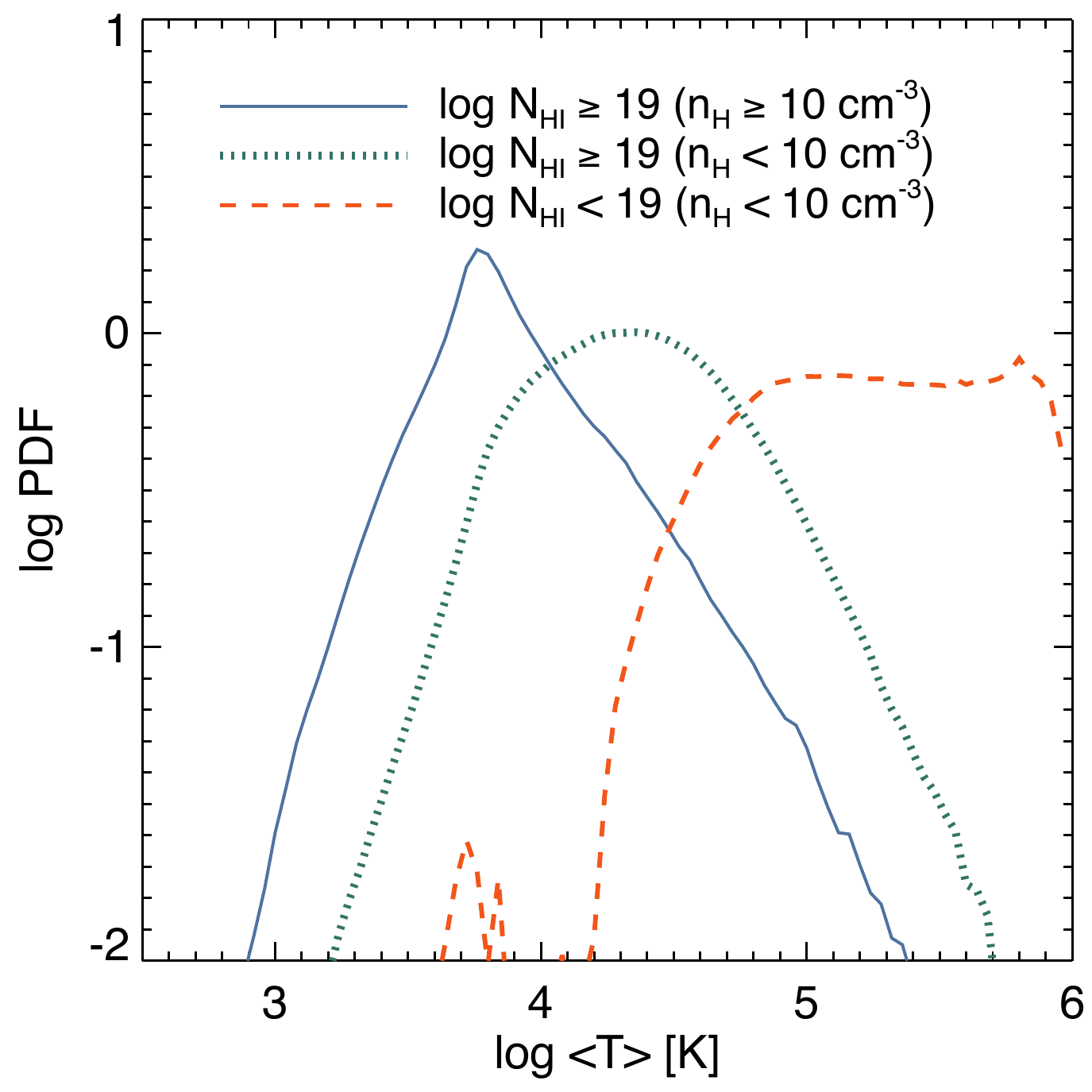}} 
\vskip -0.0cm
\caption{
{\color{burntorange}\bf Left panel:}
shows the PDFs of 
metallicity weighted total hydrogen column density, $(N_{H}/10^{21}{\rm cm}^{-2})(Z/\zsun$,
excluding gas with temperature greater than $10^6$K, 
for the three sub-populations of GRBs:
GRBs in $(n_H,T)=(10^{-2.5}{\rm cm}^{-3},10^{7.5}{\rm K})$ with $N_{\rm HI}\le 10^{19}$cm$^{-2}$ (red dashed),
$(n_H,T)=(10^{-2.5}{\rm cm}^{-3},10^{7.5}{\rm K})$ with $N_{\rm HI}\ge 10^{19}$cm$^{-2}$ (green dotted),
and 
$(n_H,T)=(10^{4.0}{\rm cm}^{-3},10^{3.75}{\rm K})$ with $N_{\rm HI}\ge 10^{19}$cm$^{-2}$ (blue solid).
The exclusion of $\ge 10^6$K gas is to intended for the situation that dust is efficiently destroyed in hot gas.
According to \citet[][]{2003Draine}, $A_V\approx (N_{H}/10^{21}{\rm cm}^{-2})(Z/\zsun$.
{\color{burntorange}\bf Right panel:}
shows the PDFs of 
gas temperature weighted by $N_{H}Z$, excluding gas with temperature greater than $10^6$K. 
}
\label{fig:Av}
\end{figure}

We now turn to the issue of dust obscuration.
The left panel of Figure~\ref{fig:Av} shows the PDFs of 
visual extinction $A_V$.
It is noted that the simulation does not follow dust formation explicitly.
Thus, we have adopted the well known empirical relation between
metal column density and visual extinction: $A_V=(N_{H}/10^{21}{\rm cm}^{-2})(Z/\zsun)$ \citep[][]{2003Draine}. 
While the applicability of the relation derived from local observations is uncertain, 
detailed analysis of galaxy colors at EoR suggest that
the simulated galaxies based on this relations give rise to self-consistent results when
comparing to observations \citep[][]{2013Kimm, 2014bCen}.
Moreover, direct observations of dust suggest that this relation holds well
in other galaxies locally, and galaxies and damped Lyman alpha systems 
at moderate to high redshift \citep[e.g.,][]{2007Draine,2013DeCia, 2014Draine, 2014Fisher}.
Nevertheless, it is possible that the normalization factor in front of the relation 
is probably uncertain to order of unity.
It is evident that a significant fraction of GRBs in the high HnHL group (blue solid curve)
are heavily dust obscured, with (53\%,16\%) of GRBs in the HnHL group have $A_V\ge (1,10)$.
At the other extreme, we see that the GRBs in the LnLN group (red dashed curve)
have negligible dust columns with no case of $A_V\ge 0.3$;
nevertheless, it is worth pointing out that, even for this set of GRBs 
12\% has an $A_V\ge 0.03$ due largely to dust in high temperature gas.
The GRBs in the LnHN group (green dotted curve) 
situates inbetween the above two groups, with a small but non-negligible fraction (7\%) at $A_V>1$.
Observationally, the issue of dust in high-z GRB hosts is less than settled.
\citet[][]{2010Zafar}, based on a re-analysis of the multi-epoch data of the afterglow of GRB 050904 at $z=6.3$, 
conclude that there is no evidence of dust.
Given that the neutral column density of the in situ DLA for GRB 050904 
is $N_{HI}=10^{21.6}$cm$^{-2}$ and low metallicity $Z=-2.6$ to $-1$
\citep[][]{2006Totani, 2006Kawai},
an $A_V\sim 0.01$ is possible, if one adopts the lower metallicity value that 
is statistically allowed in our model (see the right panel of Figure~\ref{fig:NT}).
Thus, both the low extinction and a standard ratio of dust to metals are still consistent with the observations.
Our model indicates that 9\% of GRBs have $A_V\ge 1$.
Thus, with a $z\ge 6$ GRB sample of size $11$, one expects to see one GRBs with $n_H\sim 10^{4}{\rm cm}^{-3}$ that
is significantly obscured by dust with $A_V>1$.
This may be testable with SWIFT data relatively soon.

The right panel of Figure~\ref{fig:Av} shows the PDFs of 
average gas temperature weighted by $N_{H}Z$, excluding gas with temperature greater than $10^6$K
(the exclusion is a crude way to say that dust in gas hotter than $10^6$K is destroyed). 
The purpose of this plot is to provide an indication the diversity of intervening gas with dust.
One notes that the lines of sight of HnHL GRBs contain dust in cold medium ($T\le 10^4$K),
whereas those of high LnHN GRBs are dominated by dust residing in gas $T\sim 10^{4-5}$K,
and the LnLN GRBs are intervened by dust in hotter gas of $T\ge 10^{5}$K.
Under the assumption that the hotter gas is presumably produced by shocks, which are more destructive to larger dust grains,
one might suggest that dust becomes increasingly grayer from HnHL to LnHN to LnLN. 
One expectation is that some lines of sight, especially those for the HnLN and HnHN groups 
the total dust arise from multiple, different temperature regions.
This may provide a physical explanation 
for the observational indications of multiple dust components \citep[e.g.,][]{2012Zafar}.

As a side note on the SFR of GRB hosts.
\citet[][]{2012Basa} place the star formation rate (SFR) of 
GRB 080913 at $z=6.7$ to be less than $0.9\msun$/yr.
\citet[][]{2007Berger} obtain an upper bound on the star formation rate of 
GRB 050904 at $z=6.3$ less than $5.7\msun$/yr.
From the simulations we find that 
(42\%, 57\%, 66\%) of GRBs occur in galaxies with SFR less than (0.3, 1.0, 3.0)$\msun$/yr.
Thus, while simulations and observations are in good agreement,
larger data sets are needed to place the comparisons on a solid statistical ground.

Finally, we must stress that the analysis performed here has focused on the ISM embedding the GRBs at EoR.
The exact details of the state of the IGM at EoR are uncertain both observationally and theoretically.
The theoretical difficulty is in part computational, because we do not have the capability to 
simulate a large enough volume to capture of the reionization of the IGM self-consistently, while
still having 
enough resolution for the ISM. The goal of this work is to present the signatures of
the ISM theoretically, which is lacking. It may be argued that the properties of ISM
in galaxies are somewhat detached from the properties of the IGM on large scales; in
other words, the observed spectra of GRB afterglows at EoR may be considered to be
imprinted by both ISM and IGM as a linear superposition. Consequently, a proper
understanding of the ISM will not only aid in the interpretation of the ISM of galaxies
at EoR but also is highly needed for proper interpretation of the properties (neutral
fraction, topology, etc) of the IGM at EoR.

\section{Conclusions}

We perform an analysis 
to quantify the physical condition of the ISM embedding GRBs as well as intervening gas in the host galaxies
at high redshift $z\ge 6$.
Our analysis is based on a zoomed-in cosmological radiation hydrodynamics simulation of
$3.8\times4.8\times9.6$ {\rm Mpc}$^3$ box (comoving) with 
731 halos of mass $10^{8}\le M_{\rm vir} < 3\times10^{10}\,M_{\odot}$ at $z=7$
at high spatial ($\sim$ 4 pc, physical) and stellar mass resolution of 49 $\msun$.
The following are new findings.

On parsec scales, GRBs are concentrated in two regions in 
density-temperature-metallicity space, with $(n_H,T,Z)$ being
$(10^{-2.5}{\rm cm}^{-3},10^{7.5}{\rm K}, -1.5\ {\rm to}\ 0.5)$ 
and $(>10^{4.0}{\rm cm}^{-3},<10^{3.8}{\rm K},-2.5\ {\rm to} -1)$,
consisting of 87\% and 13\% of GRBs, respectively. 
The appearance of GRB afterglows, however, also strongly depends on the line of sight
thanks to varying physical properties of intervening gas and dust in the host galaxy,
which in turn splits the low density peak into two subsets.
As a result, three separate apparent groups of GRB afterglows composing
of (13\%,37\%,50\%) are expected to arise with the following physical properties:
(1) a cold neutral circumburst gas of hydrogen density of $n_H\sim 10^4$cm$^{-3}$, 
neutral hydrogen column density 90\% range of $N_{\rm HI}\sim 10^{21.0}-10^{23.3}$cm$^{-2}$,
total hydrogen column density 90\% range of $N_{\rm H}\sim 10^{21.6}-10^{23.8}$cm$^{-2}$, 
with (53\%,16\%) having $A_V$ greater than (1,10), 
(2) a hot circumburst gas of hydrogen density of $n_H\sim 10^{-2.5}$cm$^{-3}$, 
neutral hydrogen column density 90\% range of $N_{\rm HI}\sim 10^{19.5}-10^{22.3}$cm$^{-2}$,
total hydrogen column density 90\% range of $N_{\rm H}\sim 10^{20.6}-10^{22.5}$cm$^{-2}$, 
with (7\%,1\%) having $A_V$ greater than (1,10), 
(3) a hot circumburst gas of hydrogen density of $n_H\sim 10^{-2.5}$cm$^{-3}$, 
neutral hydrogen column density 90\% range of $N_{\rm HI}\sim 10^{10.5}-10^{18.0}$cm$^{-2}$,
total hydrogen column density 90\% range of $N_{\rm H}\sim 10^{18.7}-10^{21.1}$cm$^{-2}$, 
with (0.2\%,0\%) having $A_V$ greater than (1,10). 
Common among all three groups of GRBs is that the metallicity of proximity optical/UV absorbers in the afterglow spectra
is expected to be in the range of $Z=-3$ to $-1$.

The strikingly diverse physical properties - metallicity, neutral hydrogen column density, total
column density, gas temperature, dust column - of intervening absorbers of GRB afterglows as well as
the bimodal physical properties of local (parsec-scale) environment indicates that a solid statistical
comparison between the model predictions and observations needs to await a large observational
sample of GRBs at the EoR. It is obvious that the available small sample of GRB afterglows
complicates the task of interpretating the ISM of a small number of GRB host galaxies and the
additional task of inferring the state of the IGM at EoR. Nevertheless, utilizing high-z GRBs
to probe interstellar medium and intergalactic medium must be undertaken properly taking into
consideration the physical diversities of the interstellar medium. The analysis presented will provide
a physical framework to be confronted by future observations statistically.

\citet[][]{2006Fruchter} find that GRBs 
are more concentrated in the very brightest regions of their host galaxies than are the core-collapse supernovae.
Our analysis so far has assumed that the two populations are just proportional to one another.
One may turn the argument around and use 
the observed distributions of GRBs that depend on both embedding environment and intervening material
to test the connection between GRBs and star formation.
Needless to say, larger samples will be able to shed light 
on this extremely important issue, which will have strong bearings
on relating GRB rates to cosmological reionization.

\section*{Acknowledgements}
We would like to thank Omer Bromberg for discussion.
Computing resources were in part provided by the NASA High-
End Computing (HEC) Program through the NASA Advanced
Supercomputing (NAS) Division at Ames Research Center.
The research is supported in part by NSF grant AST-1108700 and NASA grant NNX12AF91G.


\end{document}